\documentclass[journal]{IEEEtran}

\usepackage{graphicx,amssymb,lineno}
\usepackage{amsmath,amsfonts,amssymb}

\usepackage{algorithm}
\usepackage{algorithmic}
\usepackage[usenames]{color}
\usepackage{float}
\usepackage{multirow}
\usepackage{graphicx}
\usepackage{graphicx,graphics,color,epsfig,subfigure,graphpap,rotate}
\usepackage{times, verbatim, subfigure, epsfig, graphicx, latexsym, amsmath}
\usepackage{url}
\usepackage{subfigure}

\usepackage{lineno}
\usepackage{setspace}

\begin{document}
\begin{spacing}{1.0}


\title{A Survey of Millimeter Wave (mmWave) Communications for 5G: Opportunities and Challenges
}


\author{Yong~Niu,
        Yong~Li,~\IEEEmembership{Member,~IEEE,}
        Depeng~Jin,~\IEEEmembership{Member,~IEEE,}
        Li~Su,
        and Athanasios V. Vasilakos,~\IEEEmembership{Senior Member,~IEEE}
\thanks{Y. Niu, Y. Li, D.~Jin and L.~Su are with State Key Laboratory on
 Microwave and Digital Communications, Tsinghua National Laboratory for Information
 Science and Technology (TNLIST), Department of Electronic Engineering, Tsinghua
 University, Beijing 100084, China (E-mails: liyong07@tsinghua.edu.cn).} 
\thanks{A. V. Vasilakos is with the Department of Computer and Telecommunications Engineering,
 University of Western Macedonia, Greece.} %

}%

\maketitle

\begin{abstract}

With the explosive growth of mobile data demand, the fifth generation (5G) mobile network would
exploit the enormous amount of spectrum in the millimeter wave (mmWave) bands to greatly increase communication capacity. There are fundamental differences between
mmWave communications and existing other communication systems, in terms of high propagation loss,
directivity, and sensitivity to blockage. These characteristics of mmWave communications pose
several challenges to fully exploit the potential of mmWave communications, including integrated circuits and system design, interference
management, spatial reuse, anti-blockage, and dynamics control. To address these challenges, we
carry out a survey of existing solutions and standards, and propose design guidelines in
architectures and protocols for mmWave communications. We also discuss the potential applications of mmWave communications in the 5G network, including the small cell access, the cellular access, and the wireless backhaul. Finally, we discuss relevant
open research issues including the new physical layer technology, software-defined network architecture, measurements of network state information,
efficient control mechanisms, and heterogeneous networking, which should be further investigated to
facilitate the deployment of mmWave communication systems in the future 5G networks.

\end{abstract}

\section{Introduction}
\label{intro}

With the explosive growth of mobile traffic demand, the contradiction between capacity requirements
and spectrum shortage becomes increasingly prominent. The bottleneck of wireless bandwidth becomes
a key problem of the fifth generation (5G) wireless networks. On the other hand, with huge bandwidth in the millimeter wave (mmWave) band from 30 GHz to 300 GHz, millimeter wave (mmWave)
communications have been proposed to be an important part of the 5G mobile network to provide multi-gigabit communication services such as high definition television (HDTV) and ultra-high definition
video (UHDV) \cite{2_30,2_31}. Most of the current research is focused on the 28 GHz band, the 38 GHz band, the 60 GHz band, and the E-band (71--76 GHz and 81--86 GHz). Rapid progress in complementary metal-oxide-semiconductor (CMOS) radio frequency (RF)
integrated circuits paves the way for electronic products in the mmWave band \cite{CMOS,CMOS2,CMOS3}.
There are already several standards defined for indoor wireless personal area networks (WPAN) or
wireless local area networks (WLAN), for example, ECMA-387 \cite{ECMA 387,2_3} , IEEE 802.15.3c
\cite{IEEE 802.15.3c}, and IEEE 802.11ad \cite{IEEE 802.11ad}, which stimulates growing interests
in cellular systems or outdoor mesh networks in the mmWave band \cite{mmW cellular1,mmW
cellular2,mmW cellular3,mmW-cellular,singh_outdoor}.


However, due to the fundamental differences between mmWave
communications and existing other communication systems operating in
the microwaves band (e.g., 2.4 GHz and 5 GHz), there are many
challenges in physical (PHY), medium access control (MAC), and
routing layers for mmWave communications to make a big impact in the
5G wireless networks. The high propagation loss,
directivity, sensitivity to blockage, and dynamics due to mobility
of mmWave communications require new thoughts and insights in
architectures and protocols to cope
with these challenges.

In this paper, we carry out a survey of mmWave communications for 5G. We first summarize the characteristics of mmWave communications. Due to the high
carrier frequency, mmWave communications suffer from huge propagation loss, and beamforming (BF)
has been adopted as an essential technique, which indicates that mmWave communications are
inherently directional. Besides, due to weak diffraction ability, mmWave communications are
sensitive to blockage by obstacles such as humans and furniture. Then we introduce two standards
for mmWave communications in the 60 GHz band, IEEE 802.15.3c and IEEE 802.11ad. We also identify the challenges posed
by mmWave communications, and carry out a survey of existing solutions. The challenges in the integrated circuits and system design include the nonlinear distortion of power amplifiers, phase noise, IQ imbalance, highly directional antenna design, etc. Due to the directivity of
transmission, coordination mechanism becomes the key to the MAC design, and concurrent transmission
(spatial reuse) should be exploited fully to improve network capacity. To overcome blockage,
multiple approaches from the physical layer to the network layer have been proposed. However, every
approach has its advantages and shortcomings, and these approaches should be combined in an
intelligent way to achieve robust and efficient network performance. Due to human mobility and
small coverage areas of mmWave communications, dynamics in terms of channel quality and load should
be dealt with elaborately by handovers and channel state adaption mechanisms. The potential applications of mmWave communications in the 5G network include the small cell access, the cellular access, and the wireless backhaul. We then discuss some
open research issues and propose design guidelines in architectures and protocols for mmWave
communications. New physical layer technologies at mmWave frequencies including the multiple-input and multiple-output (MIMO) technique and the full-duplex technique are introduced and discussed in terms of advantages and open problems. Borrowing the idea of software defined networks \cite{SDN1}, we propose the software defined architecture for mmWave networks, and discuss the open problems therein,
such as the interface between the control plane and the data plane, centralized control mechanisms,
and network state information measurements. In the further networks, mmWave networks have to
coexist with other networks, such as LTE and WiFi. In such a heterogeneous network (HetNets), interaction and
cooperation between different kinds of networks become the key to explore the potential of
heterogeneous networking.

The rest of the paper is structured as follows. Section \ref{sec:1}
summarizes the characteristics of mmWave communications. Section
\ref{sec:2} introduces two typical standards for mmWave
communications, IEEE 802.15.3c and IEEE 802.11ad. Section
\ref{sec:3} discuss the challenges including the integrated circuits and system design, the interference management
and spatial reuse, anti-blockage, and dynamics due to user mobility. Existing solutions
for the challenges are also discussed in section \ref{sec:3}. The potential applications of mmWave communications in 5G are discussed in section \ref{sec:3-a}. In
section \ref{sec:4}, we discuss open research issues to be further
investigated. Finally, section \ref{sec:5} concludes this paper.

\section{Characteristics of mmWave communications} \label{sec:1}

The peculiar characteristics of mmWave communications should be considered in the design of network
architectures and protocols to fully exploit its potential. We summarize and present the
characteristics in the following subsections.

\subsection{Wireless Channel Measurement} \label{sec:1-a}

\begin{figure}
  \includegraphics[width=0.5\textwidth]{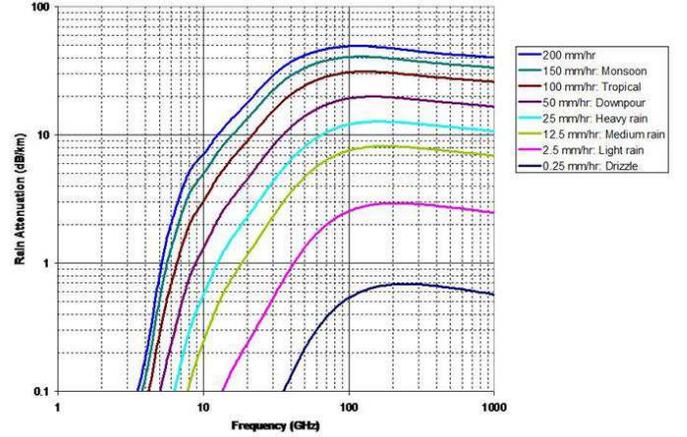}
\caption{Rain attenuation at microwave and mmWave frequencies \cite{eband-online}.}
\label{fig:rain}       
\end{figure}

\begin{figure}
  \includegraphics[width=0.5\textwidth]{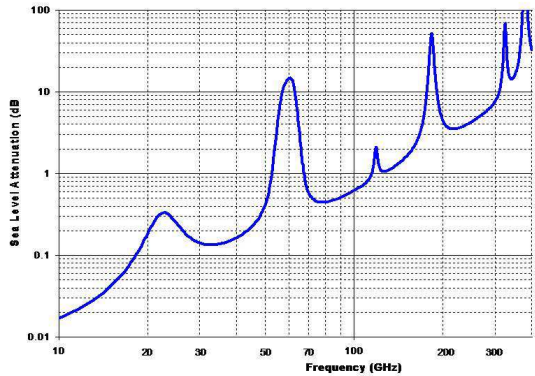}
\caption{Atmospheric and molecular absorption at mmWave frequencies \cite{eband-online}.}
\label{fig:ox}       
\end{figure}

Millimeter wave communications suffer from huge propagation loss compared with
other communication system in using lower carrier frequencies. The rain
attenuation and atmospheric and molecular absorption characteristics of
mmWave propagation limit the range of mmWave communications \cite{Rain,Rain2,eband-online}, which is shown in Fig. \ref{fig:rain} and Fig. \ref{fig:ox}. However, with smaller cell sizes applied to improve spectral efficiency today, the rain attenuation and atmospheric absorption do not create significant additional path loss for cell sizes on
the order of 200 m \cite{mmW-cellular-4}. Therefore, mmWave communications are mainly used for indoor environments, and small cell access and backhaul with cell sizes on the order of 200 m.

There have been considerable
work on mmWave propagation at the 60 GHz band \cite{CMOS3,mmW_urban,mmW channel1,mmW channel2,mmW channel3,mmW channel3-a,mmW
channel4,mmW channel5,mmW channel7,mmW channel characterization,mmW channel8}. The free space
propagation loss is proportional to the square of the carrier frequency. With a wavelength of about
5 mm, the free space propagation loss at 60 GHz is 28 decibels (dB) more than at 2.4 GHz
\cite{singh_outdoor}. Besides, the Oxygen absorption in the 60 GHz band has a peak, ranging from 15
to 30 dB/km \cite{Oxygen}. The channel characterization in \cite{NLOS} shows that the
non-line-of-sight (NLOS) channel suffers from higher attenuation than the line-of-sight (LOS)
channel. The large scale fading $F(d)$ can be modeled as follows.

\begin{equation}
F(d) = PL({d_0}) + 10n{\log _{10}}\frac{d}{{{d_0}}} - {S_\sigma },
\end{equation}
where $PL(d_0)$ is the path loss at reference distance $d_0$, $n$ is
the path loss exponent, and ${S_\sigma }$ is the showing loss.
$\sigma $ is the standard deviation of ${S_\sigma }$. In table
\ref{tab:first}, we list the statistical parameters of the path loss
model obtained in a corridor, a LOS hall, and a NLOS hall
\cite{NLOS}. We can observe that the path loss exponent in the LOS
hall is 2.17, while the path loss exponent in the NLOS hall is 3.01.
To combat severe propagation loss, directional antennas are employed
at both transmitter and receiver to achieve a high antenna gain.

\begin{table}
\centering
\caption{The statistical parameters in the path loss model.}
\label{tab:first}       
\begin{tabular}{lccc}
\hline\noalign{\smallskip}
       & $PL(d_0)$ [dB] & $n$  & $\sigma $ [dB]\\
\noalign{\smallskip}\hline\noalign{\smallskip}
Corridor & 68 & 1.64 & 2.53\\
LOS hall & 68 & 2.17 & 0.88\\
NLOS hall& 68 & 3.01 & 1.55\\
\noalign{\smallskip}\hline
\end{tabular}
\end{table}


For the small-scale propagation effects in the 60 GHz band, it is found that the
multipath effect is not obvious with directional antennas. By using
circular polarization and  receiving antennas of narrow beam width,
multipath reflection can be suppressed
\cite{Polarization1,Polarization2}. In the LOS channel model in the
conference room environment proposed in IEEE 802.11ad \cite{IEEE
802.11ad}, the direct path contains almost all the energy, and
nearly no other multipath components exist. In this case, the
channel can be regarded as the Additive White Gaussian Noise (AWGN)
channel. In the NLOS channel, there is no direct path, and the
number of paths with significant energy is small. To achieve high
data rate and maximize the power efficiency \cite{MRDMAC}, mmWave
communications mainly rely on the LOS transmission.

There are also channel measurements for mmWave cellular in other bands, such as the 28 GHz band, the 38 GHz band, and the 73 GHz band \cite{r_add1,r_add2}. Rappaport \emph{et al.} \cite{mmW-cellular-4} conducted the 28 GHz urban propagation campaign in New York City, where the distance between the transmitter (TX)
and the receiver (RX) ranged from 75 m to 125 m. The results show that the LOS path loss exponent is 2.55 resulting from all the measurements acquired in New York City. The average path loss exponent in the NLOS case is 5.76. They also conducted an outage study in Manhattan, New York \cite{outage_measure}. It was found that signal acquired by the RX for all cases was within 200 meters, and 57 \% of locations were outage due to obstruction with most of the outages beyond 200 meters from the TX. The maximum coverage distance was shown to increase with increasing antenna gains and a decrease of the path loss exponent. The maximum coverage distance achieves 200 m in a highly obstructed environment when the combined TX-RX antenna gain is 49 dBi. Zhao \emph{et al.} \cite{reflection_meas} conducted penetration and reflection measurements at
28 GHz in New York City, and it was found that tinted glass and brick pillars have high penetration losses of 40.1 dB and 28.3 dB, respectively. For indoor materials such as clear non-tinted glass and drywall, the losses are relatively low, 3.6 dB and 6.8 dB, respectively. For the reflection measurement, the outdoor materials have larger reflection coefficients, and the indoor materials have lower reflection coefficients. Samimi \emph{et al.} \cite{AOA} conducted the angle of arrival and angle of departure measurement in outdoor urban environments in New York City. It was found that New York City has rich multipath when using highly directional steerable horn antennas, and at any receiver location, there is an average of 2.5
signal lobes.

Akdeniz \emph{et al.} \cite{r_3} derived detailed spatial statistical models of channels at 28 and 73 GHz in New York, NY, USA, and the channel parameters include the path
loss, number of spatial clusters, angular dispersion, and outage. It was found that even in in highly NLOS environments, strong signals can be detected 100--200 m from potential cell sites, and spatial multiplexing and diversity can be supported at
many locations with multiple path clusters received. Nguyen \emph{et al.} \cite{r_4} conducted a wideband propagation measurement
campaign using rotating directional antennas at 73 GHz at the New York University (NYU) campus, and based on these results, they presented an
empirical ray-tracing model to predict the propagation characteristics at 73 GHz E-Band. Based on the measurement results, Thomas \emph{et al.} \cite{r_2,r_2a} developed a preliminary
3GPP-style 3D mmWave channel model by using the ray tracer to determine elevation
model parameters. Rappaport \emph{et al.} \cite{38Ghz,38Ghz-2,r_7} conducted the 38 GHz cellular propagations measurements in Austin, Texas at the University of Texas main
campus. The LOS path loss exponent for the 25 dBi horn antennas was measured to be 2.30, while the NLOS path loss exponent was 3.86. The root
mean squared (RMS) delay was shown to be higher with a lower antenna gain. Based on an outage study, it was found that base stations of lower heights have better close-in coverage, and most of the outages occur at locations beyond 200 m from the base stations. The results also show that AOAs occur mostly when the RX azimuth angle is between $- {20^ \circ }$ and $+ {20^ \circ }$ about the boresight of the TX azimuth angle \cite{38Ghz}.

\begin{table*}
\centering
\caption{The propagation characteristics of mmWave communications in different bands.}
\label{tab:measurement}       
\begin{tabular}{c|c|c|c|c|c}
\hline
\multirow{2}{*}{\textbf{Frequency Band}} & \multicolumn{2}{ |c| }{\textbf{PLE}} & \multicolumn{2}{ |c| }{\textbf{Rain Attenuation@200 m}} & \textbf{Oxygen Absorption}  \\\cline{2-5}
& \textbf{LOS} & \textbf{NLOS}& \textbf{5 mm/h} & \textbf{25 mm/h} & \textbf{@200 m}\\\hline
28 GHz &1.8$\sim$1.9 & 4.5$\sim$4.6& 0.18 dB & 0.9 dB & 0.04 dB \\
38 GHz &1.9 $\sim$2.0& 2.7$\sim$3.8& 0.26 dB & 1.4 dB & 0.03 dB  \\
60 GHz & 2.23 & 4.19 & 0.44 dB & 2 dB & 3.2 dB \\
73 GHz & 2&  2.45$\sim$2.69 & 0.6 dB & 2.4 dB & 0.09 dB\\
\hline
\end{tabular}
\end{table*}

In Table \ref{tab:measurement}, we summarize the propagation characteristics of mmWave communications in different bands in terms of the path loss exponent (PLE) under LOS and NLOS channels, the rain attenuation at 200 m, and the oxygen absorption at 200 m. We can observe that at the range of 200 m, the 28 GHz and 38 GHz bands suffer from low rain attenuation and oxygen absorption, while the rain attenuation and oxygen absorption in the 60 GHz and 73 GHz bands are significant. We can also observe that the NLOS transmission has additional propagation loss compared with the LOS transmission in the four bands.

\subsection{Directivity} \label{sec:1-b}

MmWave links are inherently directional. With a small wavelength,
electronically steerable antenna arrays can be realized as patterns
of metal on circuit board \cite{CMOS2,directional antenna
1,directional antenna 2}. Then by controlling the phase of the
signal transmitted by each antenna element, the antenna array steers
its beam towards any direction electronically and to achieve a high
gain at this direction, while offering a very low gain in all other
directions. To make the transmitter and receiver direct their beams
towards each other, the procedure of beam training is needed, and
several beam training algorithms have been proposed to reduce the
required beam training time \cite{beam_training,Beamtraining2,2_14}.

\subsection{Sensitivity to Blockage} \label{sec:1-c}

Electromagnetic waves have weak ability to diffract around obstacles
with a size significantly larger than the wavelength. With a small
wavelength, links in the 60 GHz band are sensitive to blockage by
obstacles (e.g., humans and furniture). For example, blockage by a
human penalizes the link budget by 20-30 dB \cite{MRDMAC}. Collonge
\emph{et al.} \cite{blockage rate} conducted propagation
measurements in a realistic indoor environment in the presence of
human activity, and the results show that the channel is blocked for
about 1\% or 2\% of the time for one to five persons. Taking human
mobility into consideration, mmWave links are intermittent.
Therefore, maintaining a reliable connection for delay-sensitive
applications such as HDTV is a big challenge for mmWave
communications.

\section{Standardization}\label{sec:2}
 Due to the great potential of mmWave communications, multiple
 international organizations have emerged for the standardization,
 including ECMA \cite{ECMA 387}, IEEE 802.15.3 Task Group 3c (TG3c) \cite{IEEE 802.15.3c}, IEEE 802.11ad
standardization task group \cite{IEEE 802.11ad}, the WirelessHD
consortium \cite{WirelessHD}, and the Wireless Gigabit Alliance
(WiGig) \cite{Wireless Gigabit Alliance}. We review two standards in
this section, IEEE 802.11ad and IEEE 802.15.3c, and further discuss
the challenges in the next section.

\subsection{IEEE 802.11ad}

IEEE 802.11ad specifies the physical layer and MAC layer in 60GHz
band to support multi-gigabit wireless applications including
instant wireless sync, wireless display of high definition (HD)
streams, cordless computing, and internet access \cite{IEEE
802.11ad}. In the physical layer, two operating modes are defined,
the orthogonal frequency division multiplexing (OFDM) mode for high
performance applications (e.g. high data rate), and the single
carrier (SC) mode for low power and low complexity implementation.


In IEEE 802.11ad, a basic service set (BSS) consists of a designated device, called AP, and N
non-AP devices (DEVs). AP provides the basic timing for the BSS, and coordinates medium access in
the BSS to accommodate traffic requests from the DEVs. The channel access time is divided into a
sequence of beacon intervals (BIs), and each BI consists of four portions including beacon
transmission interval (BTI), association BF training (A-BFT), announcement transmission interval
(ATI), and data transfer interval (DTI). In BTI, AP transmits one or more mmWave beacon frames in a
transmit sector sweep manner. Then initial BF training between AP and non-AP DEVs, and association
are performed in A-BFT. Contention-based access periods (CBAPs) and service periods (SPs) are
allocated within each DTI by AP during ATI. During DTI, peer-to-peer communications between any
pair of DEVs including the AP and the non-AP DEVs are supported after completing the beamforming
(BF) training. In IEEE 802.11ad, a hybrid multiple access of carrier sensing multiple
access/collision avoidance (CSMA/CA) and time division multiple access (TDMA) is adopted for
transmissions among devices. CSMA/CA is more suitable for bursty traffic such as web browsing to
reduce latency, while TDMA is more suitable for traffic such as video transmission to support
better quality of service (QoS).


\subsection{IEEE 802.15.3c}

IEEE 802.15.3c specifies the physical layer and MAC layer for indoor
WPANs (also referred to as the piconet) composed of several wireless
nodes (WNs) and a single piconet controller (PNC). The PNC provides
network synchronization and coordinates the transmission in the
piconet. In IEEE 802.15.3c, network time is divided into a sequence
of superframes, each of which consists of three portions: the beacon
period (BP), the contention access period (CAP), and the channel
time allocation period (CTAP). During BP, network synchronization
and control messages are broadcasted from the PNC. CAP is for
devices to send transmission requests to the PNC by the CSMA/CA
access method, and CTAP is for data transmissions among devices.
During CTAP, TDMA is applied, and each time slot is scheduled to a
specific flow.

\section{Challenges and Existing Solutions}\label{sec:3}

Despite the potential of mmWave communications, there are a number
of key challenges to exploit the benefits of mmWave communications. Now, we discuss the challenges and present
related existing solutions.

\subsection{Integrated Circuits and System Design}

With high carrier frequency and wide bandwidth, there are several technical challenges in the design of circuit
components and antennas for mmWave communications \cite{CMOS3}. In the 60 GHz band, high transmit power, i.e., equivalent
isotropic radiated power (EIRP), and huge bandwidth cause severe nonlinear distortion of power amplifiers (PA) \cite{PA}. Besides, phase noise and IQ imbalance are also challenging problems faced by radio frequency (RF) integrated circuits \cite{PA,IQ}.

Research progress on integrated circuits for mmWave communications in the 60 GHz band has been summarized and discussed in \cite{CMOS3}, including on-chip and in-package antennas, radiofrequency (RF) power amplifiers (PAs), low-noise amplifiers (LNAs), voltage-controlled oscillators (VCOs), mixers, and analog-to-digital converters (ADCs). Hong \emph{et al.} \cite{2_1} deduced a novel and practical phased array antenna solution operating at 28 GHz with near spherical coverage. They also designed cellular phone prototype equipped with mmWave 5G antenna arrays consisting of a total of 32 low-profile antenna elements. Hu \emph{et al.} \cite{2_2} presented a cavity-backed slot (CBS) antenna for millimeter-wave applications. The cavity of the antenna is fully filled by polymer material, which reduces the cavity size by 76.8 \%. Liao \emph{et al.} \cite{2_12} presented a novel planar aperture antenna with differential feeding, which maintains a high gain and wide bandwidth compared with conventional high gain aperture antennas. The proposed aperture antenna element has low cost, low profile, compact size, and is also good in gain and bandwidth. Zwick \emph{et al.} \cite{2_15} presented a new planar superstrate antenna suitable for integration with mmWave transceiver integrated circuits, which is printed on the bottom of a dielectric superstrate with a ground plane below. Two designs for the 60 GHz band achieve over 10\% bandwidth while maintaining better than 80\% efficiency.

\subsection{Interference Management and Spatial Reuse}

The directivity of transmission enables less interference between
links. In the outdoor mesh network in the 60 GHz band, the highly
directional links are modeled as \emph{pseudowired}, and the
interference between nonadjacent links is negligible
\cite{pseudo-wired}. The details of antenna patterns can also be
ignored in the design of MAC protocols for mmWave mesh networks. Due
to directional transmission, the third party nodes cannot perform
carrier sense as in WiFi, which is referred to as the deafness
problem \cite{mao}. In this case, the coordination mechanism becomes
the key to the MAC design, and concurrent transmission should be
exploited fully to greatly enhance the network capacity
\cite{singh_outdoor}.

\begin{figure}
  \includegraphics[width=0.45\textwidth]{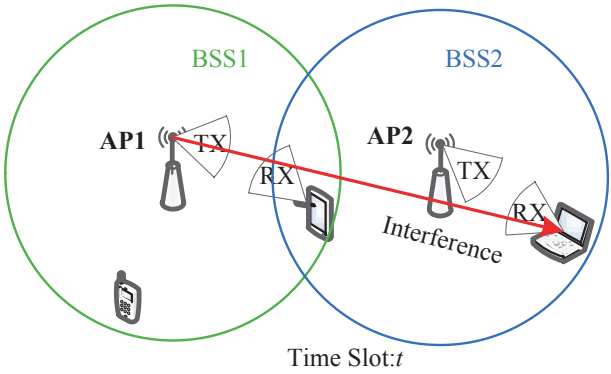}
\caption{Interference among different BSSs}
\label{fig:2}       
\end{figure}

In the indoor environments, however, due to the limited range, the
assumption of \emph{pseudowired} is not reasonable
\cite{Qiao,Qiao_6}. On the other hand, owing to the explosive growth
of mobile data demands as well as to overcome the limited range of
mmWave communications, in a practical mmWave communication system,
the number of deployed APs over both public and private areas
increases tremendously. For example, a large number of APs must be
deployed in scenarios such as enterprise cubicles and conference
rooms to provide seamless coverage. In this case, the interference
in the network can be divided into two portions: interference within
each BSS, and interference among different BSSs \cite{obss}. As
shown in Fig. \ref{fig:2}, when the two links in BSS1 and BSS2 are
communicating in the same slot $t$, since AP1 directs its beam
towards the laptop, AP1 will have interference to the laptop. If the
distance between them is short, the service of the laptop will be
degraded significantly.

Thus, interference management mechanisms such as power control and
transmission coordination should be applied to avoid significant
degradation of network performance due to interference. With the
interference efficiently managed, concurrent transmission (spatial
reuse) on the other hand should be supported among different BSSs as
well as within each BSS \cite{mao,Qiao,chen_2}.

In order to address these problems, there has been some related work on directional MAC protocols
for mmWave communications. Since TDMA is adopted in ECMA-387 \cite{ECMA 387}, IEEE 802.15.3c
\cite{IEEE 802.15.3c}, and IEEE 802.11ad \cite{IEEE 802.11ad}, many protocols are based on TDMA
\cite{mao_12,mao_13}. Cai \emph{et al.} \cite{EX Region} introduced the concept of exclusive region
(ER) to enable concurrent transmissions, and derived the ER conditions that concurrent
transmissions always outperform TDMA for both omni-antenna and directional-antenna models. By the
REX scheduling scheme (REX), significant spatial reuse gain is achieved. However, the interference
level and the received signal power are calculated by the free space path loss model, which is
inappropriate for indoor WPANs, where reflection will also cause interference. Besides, it only
considers the two-dimensional space in the transmission scheduling problem, and the power control
is not considered to manage interference. In two protocols based on IEEE 802.15.3c, multiple links
are scheduled to communicate in the same slot if the multi-user interference (MUI) is below a
specific threshold \cite{Qiao_6,Qiao_15}. However, they do not capture the characteristics of the
directional antennas, and the aggregation effect of interference from multiple links is also not
considered. Qiao \emph{et al.} \cite{Qiao} proposed a concurrent transmission scheduling algorithm
for an indoor IEEE 802.15.3c WPAN, and non-interfering and interfering links are scheduled to
transmit concurrently to maximize the number of flows with the quality of service requirement of
each flow satisfied. It can support more number of users, and significantly improves the resource
utilization efficiency in mmWave WPANs. However, it does not consider the NLOS transmissions, which
is important for indoor WPANs, and the interference model does not take the realistic antenna model
into account. Furthermore, a multi-hop concurrent transmission scheme (MHCT) is proposed to address
the link outage problem (blockage) and combat huge path loss to improve flow throughput
\cite{Qiao_7}. It assumes an ideal ``flat-top'' model for directional antenna, and analyzes the
spatial reuse and time division multiplexing gain of MHCT in the two-dimensional space. Based on
IEEE 802.15.3c, the piconet controller selects appropriate relay hops for a traffic flow according
to a hop selection metric, and spatial reuse is also exploited by the multi-hop concurrent
transmission scheme (MHCT). For protocols based on IEEE 802.15.3c, the piconet controller is
operating in the omni-directional mode during the random access period to avoid the deafness
problem, which may not be feasible for mmWave systems that operate in the multi-gigabit domain with
highly directional transmission, and will lead to the asymmetry-in-gain problem \cite{DtDMAC}. For
TDMA based protocols, the medium time for bursty data traffic is often highly unpredictable, which
will cause some flows to have too much medium time while not enough medium time for others.
Besides, the control overhead for on-the-fly medium reservation may be high for TDMA based
protocols. Based on IEEE 802.11 ad, Chen \emph{et al.} \cite{chen_2} proposed a spatial reuse
strategy to schedule two different SPs to overlap with each other, and also analyzed the
performance of the strategy with the difference between idealistic and realistic directional
antennas considered. It does not fully exploit the spatial reuse since only two links are
considered for concurrent transmissions.


On the other hand, some protocols are based on the centralized coordination by the AP or PNC. Gong
\emph{et al.} \cite{Gong} proposed a directive CSMA/CA protocol, which exploits the virtual carrier
sensing to solve the deafness problem completely. The network allocation vector (NAV) information
is distributed by the PNC. Spatial reuse, however, is not fully exploited to improve network
capacity in the protocol. Son \emph{et al.} \cite{mao} proposed a frame based directive MAC
protocol (FDMAC). The high efficiency of FDMAC is achieved by amortizing the scheduling overhead
over multiple concurrent transmissions in a row. The core of FDMAC is the Greedy Coloring
algorithm, which fully exploits spatial reuse and greatly improves the network throughput compared
with MRDMAC \cite{MRDMAC} and memory-guided directional MAC (MDMAC) \cite{MDMAC}. FDMAC also has a
good fairness performance and low complexity. FDMAC, however, assumes the \emph{pseudowired}
interference model for WPANs, which is not reasonable due to the limited range. Chen \emph{et al.}
\cite{chenqian} proposed a directional cooperative MAC protocol (D-CoopMAC) to coordinate the
uplink channel access among stations in an IEEE 802.11ad WLAN. In D-CoopMAC, a two-hop path of high
channel quality from the source station (STA) to the destination station (STA) is established to
replace the direct path of poor channel quality. By the two-hop relaying, D-CoopMAC significantly
improves the system throughput. However, spatial reuse is also not considered in D-CoopMAC since
most transmissions go through the AP. Park \emph{et al.} \cite{2_4} proposed an incremental multicast grouping (IMG) scheme to maximize the sum rate of devices, where adaptive beamwidths are generated depending on the locations of multicast devices. Simulation based on the IEEE 802.11ad demonstrate that the IMG scheme can improve the overall throughput by 28 \% to 79 \% compared with the conventional multicast schemes. Scott-Hayward \emph{et al.} \cite{2_8} proposed to use the particle swarm optimization (PSO) for the channel-time allocation of a mixed set of
multimedia applications in IEEE 802.11ad. Channel-time allocation PSO (CTA-PSO) is demonstrated to allocate resource successfully even when blockage occurs.

For outdoor mesh networks in the 60 GHz band, Singh \emph{et al.} \cite{MDMAC} proposed a
distributed MAC protocol, the memory-guided directional MAC (MDMAC), based on the
\emph{pseudowired} link abstractions. A Markov state transition diagram is incorporated into the
protocol to alleviate the deafness problem. MDMAC employs memory to achieve approximate time
division multiplexed (TDM) schedules, and does not fully exploit the potential of spatial reuse.
Another distributed MAC protocol for directional mmWave networks is directional-to-directional MAC
(DtDMAC), where both senders and receivers operate in a directional-only mode \cite{DtDMAC}, which
solves the asymmetry-in-gain problem. DtDMAC adopts an exponential backoff procedure for
asynchronous operation, and the deafness problem is also alleviated by a Markov state transition
diagram. DtDMAC is fully distributed, and does not require synchronization. However, it does not
capture the characteristics of wireless channel in mmWave bands, and only gives the analytical
network throughput of DtDMAC for the mmWave technology.


\subsection{Anti-blockage}

Sato and Manabe \cite{path visibility} estimated the
propagation-path visibility between APs and terminals in office
environments where blockage by human bodies happens. To avoid
shadowing by human bodies perfectly without multi-AP diversity, a
lot of APs are needed. However, diversity switching between only two
APs provides 98\% propagation path visibility. Dong \emph{et al.}
\cite{link_blockage_analysis} analyzed the link blockage probability
in typical indoor environments under random human activities. The AP
is mounted on the ceiling, and this work mainly focuses on the links
between the AP and the user devices. The results show that as the
user devices move towards the edge of the service area, the blockage
probability of links increases almost linearly. With communications
between user devices enabled, the blockage probability of links
between user devices should also be considered.

To ensure robust network connectivity, different approaches from the physical layer to the network
layer have been proposed. Genc \emph{et al.} \cite{robust} exploited reflections from walls and
other surfaces to steer around the obstacles. Yiu \emph{et al.} \cite{2_6} used static
reflectors to maintain the coverage in the entire room when blockage occurs. Using reflections will cause additional power loss
and reduce power efficiency. Besides, the node placement and environment will have a big impact on
the efficacy of reflection to overcome blockage. An \emph{et al.} \cite{beam switching} resolved
link blockage by switching the beam path from a LOS link to a NLOS link. NLOS transmissions suffer
from significant attenuation and cannot support high data rate \cite{NLOS,MRDMAC,Qiao_7}. Park and
Pan \cite{EG} proposed a spatial diversity technique, called equal-gain (EG) diversity scheme,
where multiple beams along the $N$ strongest propagation paths are formed simultaneously during a
beamforming process. When the strongest path is blocked by obstacles, the remaining paths can be
used to maintain reliable network connectivity. This approach adds the complexity and overhead of
the beamforming process, which will degrade the system performance eventually. Xiao \cite{Xiao}
proposed a suboptimal spatial diversity scheme called maximal selection (MS) by tracing the
shadowing process, which outperforms EG in terms of link margin and saves computation complexity.
Another approach is to use relays to maintain the connectivity \cite{MRDMAC,blockage and directivity
conference}. The multihop relay directional MAC (MRDMAC) overcomes the deafness problem by
PNC's weighted round robin scheduling. In MRDMAC \cite{MRDMAC}, if a wireless terminal (WT) is lost
due to blockage, the access point (AP) will choose a WT among the live WTs as a relay to the lost
node. By the multi-hop MAC architecture, MRDMAC is able to provide robust connectivity in typical
office settings. Since most transmissions go through the PNC, concurrent transmission is also not
considered in MRDMAC. Based on IEEE 802.15.3, Lan \emph{et al.} \cite{Lan} exploited the two-hop
relaying to provide alternative communication links under such harsh environments. The transmission
from relay to destination of one links is scheduled to coexist with the transmission from source to
relay of another link to improve throughput and delay performance. However, only two links are
scheduled for concurrent transmissions in this scheme, and the spatial reuse is not fully
exploited. Lan \emph{et al.} \cite{2_11} also proposed a deflection routing scheme to improve
the effective throughput by sharing time slots for direct path
with relay path. It includes a routing algorithm, the best fit deflection routing (BFDR), to find the relay path with the least interference that
maximizes the system throughput. They also developed the sub-optimal random fit deflection routing (RFDR), which achieves almost the
same order of throughput improvement with much lower complexity. With multiple APs deployed, handovers can be performed between APs to address the
blockage problem. Zhang \emph{et al.} \cite{multiap} took advantage of multi-AP diversity to
overcome blockage. There is an access controller (AC) in the multi-AP architecture, and when one of
wireless links is blocked, another AP can be selected to complete remaining transmissions. To
ensure the efficacy of this approach, multiple APs need to be deployed, and their locations will
have a significant impact on the robustness and efficiency of this approach. Recently, Niu \emph{et
al.} \cite{tvt_own} proposed a blockage robust and efficient directional MAC protocol (BRDMAC),
which overcomes the blockage problem by two-hop relaying. In BRDMAC, relay selection and spatial
reuse are optimized jointly to achieve near-optimal network performance in terms of delay and
throughput. However, only two-hop relaying is considered in BRDMAC, and under serious blockage
conditions, there is probably no two-hop relay path between the sender and the receiver, which
cannot guarantee robust network connectivity. In the network layer, Wang \emph{et al.}
\cite{multipath} exploited multipath routing to enhance reliability of high quality video in the 60GHz
radio indoor networks. It mainly focuses on the video traffic, and other traffic patterns are not
considered.

\subsection{Dynamics due to User Mobility}

User mobility poses several challenges in the mmWave communication
system. First, user mobility will incur significant changes of the
channel state. When users move, the distance between the transmitter
(TX) and the receiver (RX) varies, and the channel state also varies
accordingly. In table \ref{tab:1}, we list the channel capacities
under difference distances between TX and RX, adopting the PHY
parameters in \cite{Bogao_para}. We assume LOS transmission between
TX and RX, and thus calculate the capacities according to Shannon's
channel capacity. We can observe that the channel capacity vary with
the distance significantly. Therefore, the selection of modulation
and coding schemes (MCS) should be performed according to the
channel states to fully exploit the potential of mmWave
communications \cite{selection of MCS}.

\begin{table*}
\centering
\caption{The channel capacities under difference distances}
\label{tab:1}       
\begin{tabular}{lcccccc}
\hline\noalign{\smallskip}
Distance (m) & 1 & 2 & 4 & 6 & 8 & 10 \\
\noalign{\smallskip}\hline\noalign{\smallskip}
Capacity (Gbps) & 16.02 & 12.51 & 9.05 & 7.08 & 5.74 & 4.75 \\
\noalign{\smallskip}\hline
\end{tabular}
\end{table*}

Second, due to the small coverage areas of BSSs, especially in indoor environments, user mobility
will cause significant and rapid load fluctuations in each BSS \cite{SoftRAN}. Thus, user
association and handovers between APs should be carried out intelligently to achieve an optimized
load balance. Current standards for mmWave communications, such as IEEE 802.11ad and IEEE
802.15.3c, adopt the received signal strength indicator (RSSI) for user association, which may lead
to inefficient use of resources \cite{RSSI1,RSSI2,RSSI3}. With load, channel quality, and the
characteristics of 60 GHz wireless channels taken into consideration, Athanasiou \emph{et al.}
\cite{User association} designed a distributed association algorithm (DAA), based on Langragian
duality theory and subgradient methods. DAA is shown to be asymptotically optimal, and outperforms
the user association policy based on RSSI in terms of fast convergence, scalability, time
efficiency, and fair execution.

Besides, user mobility will also incur frequent handovers between APs. Handover mechanisms have a
big impact on quality of service (QoS) guarantee, load balance, and network capacity, etc. For
example, smooth handovers are needed to reduce dropped connections and ping-pong (multiple
handovers between the same pair of APs). However, there is little work on the handover mechanisms
for mmWave communications in the 60 GHz band. Quang \emph{et al.} \cite{handover} discussed the
handover issues in radio over fiber network at 60 GHz, and the handover performance can be improved
using more information such as velocity and mobility direction of users. Tsagkaris \emph{et al.}
\cite{handover2} proposed a novel handover scheme based on Moving Extended Cells (MEC) \cite{MEC}
to achieve seamless broadband wireless communication.



\section{Applications of mmWave Communications} \label{sec:3-a}

\subsection{Small Cell Access}

To keep up with the explosive growth of mobile traffic demand, massive densification of small cells has been proposed to achieve the 10 000 fold increase in network capacity by 2030 \cite{2_10,2030,ultra_dense}. Small cells deployed underlaying the macrocells as WLANs or WPANs are a promising solution for the capacity enhancement in the 5G cellular networks. With huge bandwidth, mmWave small cells are able to provide the multi-gigabit rates, and wideband multimedia applications such as high-speed data transfer between devices, such as
cameras, pads, and personal computers, real-time
streaming of both compressed and uncompressed high definition
television (HDTV), wireless gigabit ethernet, and wireless gaming can be supported.

Ghosh \emph{et al.} \cite{2_10} made a case for using mmWave bands, in
particular the 28, 38, 71--76 and 81--86 GHz bands for the 5G enhanced local area (eLA)
access. With huge bandwidth, the eLA system is able to achieve peak data rates in excess of
10 Gbps and edge data rates of more than 100 Mbps. Singh \emph{et al.} \cite{2_20} presented an mmWave system for
supporting uncompressed high-definition (HD) video up to 3
Gb/s. Wu \emph{et al.} \cite{2_26} defined and evaluated
important metrics to characterize multimedia QoS, and designed a QoS-aware multimedia
scheduling scheme to achieve the trade-off
between performance and complexity.

\subsection{Cellular Access}

\begin{figure}
  \includegraphics[width=0.48\textwidth]{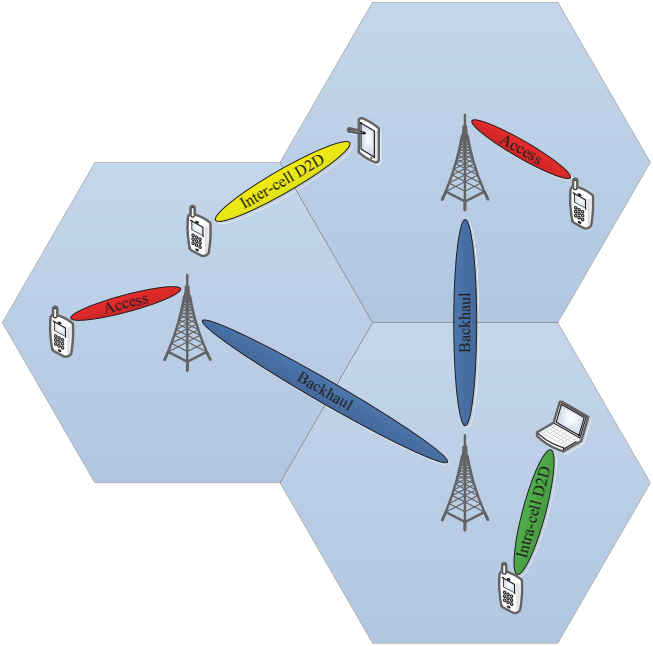}
\caption{MmWave 5G cellular network architecture with D2D communications enabled.}
\label{fig:cellular}       
\end{figure}

The large bandwidth in the mmWave bands promotes the usage of mmWave communications in the 5G cellular access \cite{mmW-cellular,mmW-cellular-4,mmW-cellular-5}. In \cite{2_19,2_21}, it is shown that mmWave cellular
networks have the potential for high coverage
and capacity as long as the infrastructure is
densely deployed. Based on the extensive propagation measurement campaigns at mmWave frequencies \cite{mmW-cellular-4}, the feasibility and efficiency of applying mmWave communications in the cellular access have been demonstrated at 28 GHz and 38 GHz with the cell sizes at the order of 200 m. It is shown in \cite{r_1} that the capacity gains based on arbitrary pointing angles of directional
antennas be 20 times greater than the 4G LTE networks, and can be further improved when directional antennas are pointed in the strongest transmit and receive directions. Since device-to-device (D2D) communications in close proximity save power and improve the spectral
efficiency, D2D communications should be enabled in the mmWave cellular systems to support the context-aware applications that involve
discovering and communicating with nearby devices. In Fig. \ref{fig:cellular}, we plot the mmWave 5G cellular network architecture with D2D communications enabled. With cellular cells densely deployed, we assume two D2D modes in the system, the intra-cell D2D transmission and the inter-cell D2D transmission. With the access link, the backhaul link, the intra-cell D2D link, and the inter-cell D2D link enabled in the mmWave band, the efficient and flexible radio resource management schemes including power control, transmission scheduling, user access, and user association, are needed to fully unleash the potential of mmWave communications.


\subsection{Wireless Backhaul}

\begin{figure}
  \includegraphics[width=0.45\textwidth]{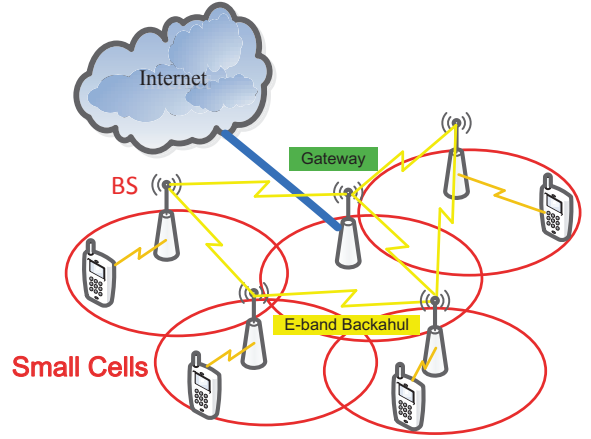}
\caption{E-band backhaul for small cells densely deployed.}
\label{fig:backhaul}       
\end{figure}

With small cells densely deployed in the next generation of cellular systems (5G), it is costly to connect 5G base stations (BSs) to the other 5G BSs and to the network by fiber based backhaul \cite{2_7}. In contrast, high speed wireless backhaul is more cost-effective,
flexible, and easier to deploy. With huge bandwidth available, wireless backhaul in mmWave bands, such as the 60 GHz band and E-band (71--76 GHz and 81--86 GHz), provides several-Gbps data rates and can be a promising backhaul solution for small cells. As shown in Fig. \ref{fig:backhaul}, the E-band backhaul provides the high speed transmission between the small cell base stations (BSs) or between BSs and the gateway.

Taori \emph{et al.} \cite{2_7} proposed to use the in-band wireless backhaul to obtain a cost-effective and scalable wireless backhaul
solution, where the backhaul and access are multiplexed on the same frequency band. They also proposed a time-division multiplexing (TDM)-based scheduling scheme to support point-to-multipoint, non-line-of-sight, mmWave backhaul. In the in-band backhaul scenario, the joint design of the access and backhaul networks will optimize the resource allocation further \cite{joint_design}. Recently, Niu \emph{et al.} \cite{JSAC_own} proposed a
joint transmission scheduling scheme for the radio access and
backhaul of small cells in 60 GHz band, termed D2DMAC, where
a path selection criterion is designed to enable D2D
transmissions for performance improvement.

\begin{table*}
\centering
\caption{Applications of mmWave communications.}
\label{tab:app}       
\begin{tabular}{l|c|c|c}
\hline
\textbf{Publication} & \textbf{Frequency Band (GHz)} & \textbf{Scenario} & \textbf{Application}  \\\hline
Singh \emph{et al.} \cite{MRDMAC} & 60 & indoor office & Internet access \\
Son \emph{et al.} \cite{mao} & 60 & WPAN & transmission between devices \\
Qiao \emph{et al.} \cite{Qiao} & 60 & WPAN& flows with QoS requirements\\
Chen \emph{et al.} \cite{chenqian} & 60 & WLAN & uplink channel access \\
Ghosh \emph{et al.} \cite{2_10} & 28, 38, 71--76, 81--86 & urban
street & access and backhaul  \\
Singh \emph{et al.} \cite{2_20} & 60 & WPAN & HD video \\
Wu \emph{et al.} \cite{2_26} & 60, 70 & indoor  & multimedia\\
Taori \emph{et al.} \cite{2_7}& 28 & outdoor cellular & in-band backhaul  \\
Niu \emph{et al.} \cite{JSAC_own} & 60 & small cells in HetNets & access, backhaul, D2D\\
Qiao \emph{et al.} \cite{2_29} & not specified & outdoor cellular & access, backhaul, D2D \\
\hline
\end{tabular}
\end{table*}

In Table \ref{tab:app}, we list the typical works according to the frequency band, the scenario, and the main application. We can observe that there are many works on the indoor WPAN/WLAN applications in the 60 GHz band, and the system design for the 5G cellular and HetNets in other bands should be investigated further.

\section{Open Research Issues}\label{sec:4}


In this section, we discuss the open research issues that need to be
investigated, and new research directions for mmWave communications
to make significant impact on the next generation wireless networks.

\subsection{New Physical Layer Technology}

\subsubsection{MIMO at mmWave Frequencies}

Since the MIMO techniques provide the trade-off between the multiplexing gain and the diversity gain, there is increasing interest in applying MIMO techniques in mmWave communications \cite{2_22,2_28,2_5}. However, the baseband precoding structure in the sub 3 GHz system cannot be applied directly into the mmWave system \cite{2_16}. On one hand, the baseband precoding requires a complete dedicated RF chain for each antenna element at the transmitter
or the receiver, which is expensive and adds the system complexity significantly. On the other hand, with highly directional beams, there is a shortage of multipath in the mmWave bands, and the diversity gain is low for the baseband precoding. To obtain the benefits of MIMO and also provide high beamforming gain to overcome high propagation loss in mmWave bands, the hybrid beamforming structure has been proposed as an
enabling technology for 5G cellular communications \cite{mmW_mimo}.

\begin{figure}
  \includegraphics[width=0.5\textwidth]{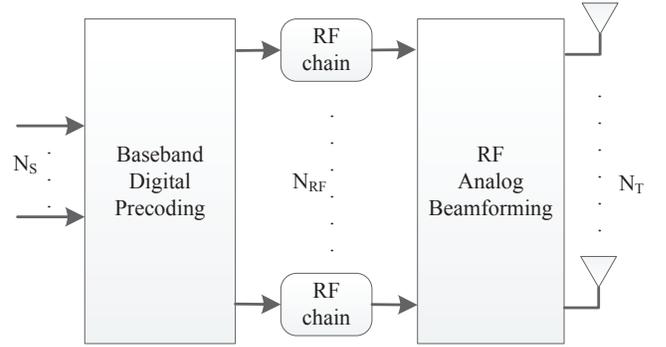}
\caption{Hybrid beamforming structure for the mmWave transmitter.}
\label{fig:mimo}       
\end{figure}

The hybrid beamforming structure for the mmWave transmitter is illustrated in Fig. \ref{fig:mimo}. In the structure, $N_S$ data streams are first through the baseband digital precoding, and then the output is through the $N_{RF}$ RF chains. After the RF analog beamforming, the RF signals are outputted to the $N_T$ antennas. With $N_T \ge N_{RF}$, the number of RF chains can be reduced for practical implementation \cite{2_16}.

Based on the hybrid beamforming structure, there has been some work on the design of the digital precoder and the analog beamformer. Ayach \emph{et al.} \cite{2_17} exploited the spatial structure of mmWave channels to formulate the transmit precoding and receiver combining problem as a sparse reconstruction problem. They also developed algorithms to accurately approximate optimal unconstrained precoders and
combiners, which can be implemented in low-cost RF hardware. Alkhateeb \emph{et al.} \cite{2_18} developed an adaptive algorithm
to estimate the mmWave channel parameters, which exploits the poor scattering nature of the channel. They also proposed a
new hybrid analog/digital precoding algorithm to overcome the hardware constraints on the analog-only beamforming, which approaches the performance of digital solutions. In \cite{2_22}, it is shown by measurements and realistic channel models that spatial multiplexing (SM) and beamforming (BF) could
work in tandem. The LOS channels are often of low rank, and only a couple of SM streams can be
supported. NLOS channels offer more rank, but they have higher path loss, which indicates both SM and BF should be exploited for capacity gain. Singh \emph{et al.} \cite{2_23} developed reduced complexity algorithms for optimizing the choice of beamforming directions of multiple antenna arrays at the transmitter or the receiver in a codebook-based beamforming
system, exploiting the sparse multipath structure of the mmWave channel. The cardinality of the joint beamforming search space is reduced by focusing on a small set of dominant candidate directions. Han \emph{et al.} \cite{2_27} investigated the optimal designs of hybrid beamforming structures, focusing on an N (the number of transceivers) by M (the number of active antennas per transceiver) hybrid beamforming structure. The energy efficiency
and spectrum efficiency of the beamforming structure are also discussed, which provides guidelines to achieve optimal energy/
spectrum efficiency trade-off. Alkhateeb \emph{et al.} \cite{2_28} reviewed two potential mmWave MIMO architectures, the hybrid analog/digital precoding/combining, and combining with low-resolution analog-to-digital converters. The advantages and disadvantages of combining with low-resolution analog-to-digital converters are discussed and analyzed. Most of the current work is focused on the single-user mmWave MIMO systems, and multi-user mmWave MIMO systems should be investigated for the mmWave cellular systems.


In Table \ref{tab:MIMO}, we summarize these works according to the transceiver structure and application scenario. We can observe that the analog
beamforming structure is mainly used in the short-range communication scenario for overcoming the path loss, while the hybrid beamforming structure is proposed
in the 5G cellular networks to achieve the trade-off between performance and complexity.

\begin{table*}
\centering
\caption{The transceiver structures for mmWave communications.}
\label{tab:MIMO}       
\begin{tabular}{l|c|c|c}
\hline
\textbf{Publication} & \textbf{Scenario} & \textbf{Structure} & \textbf{Remark}  \\\hline
Wang \emph{et al.} \cite{beam_training} & WPAN & analog beamforming & based on beam codebook \\\hline
\multirow{2}{*}{Tsang \emph{et al.} \cite{Beamtraining2}} & \multirow{2}{*}{WPAN} & \multirow{2}{*}{analog beamforming} & each beam angle is\\&&& assigned unique signature code \\\hline
\multirow{2}{*}{Alkhateeb \emph{et al.} \cite{2_28}} & \multirow{2}{*}{cellular} & \multirow{2}{*}{hybrid beamforming} & combining with low-resolution \\&&&analog-to-digital converters\\\hline
\multirow{3}{*}{Roh \emph{et al.} \cite{mmW_mimo}} & \multirow{3}{*}{cellular} & \multirow{3}{*}{hybrid beamforming} & supporting outdoor and
indoor \\&&&coverage of a few hundred meters \\&&& with more than 500 Mb/s data
rate \\\hline
\multirow{2}{*}{Ayach \emph{et al.} \cite{2_17}} & \multirow{2}{*}{cellular} & \multirow{2}{*}{hybrid beamforming} & exploit the
spatial structure\\ & & & of channels to design precoders\\\hline
\multirow{2}{*}{Alkhateeb \emph{et al.} \cite{2_18}} & \multirow{2}{*}{cellular} & \multirow{2}{*}{hybrid beamforming} & exploit the
poor scattering \\
& & & nature to estimate the channel  \\\hline
\multirow{2}{*}{Singh \emph{et al.} \cite{2_23}} & \multirow{2}{*}{cellular} & \multirow{2}{*}{hybrid beamforming} & with multiple arrays beamforming \\&&&independently, and codebook-based \\\hline
\multirow{2}{*}{Han \emph{et al.} \cite{2_27}} & \multirow{2}{*}{cellular} & \multirow{2}{*}{hybrid beamforming} & each of the $N$ transceivers \\&&&is connected
to $M$ antennas.\\\hline
\end{tabular}
\end{table*}

\subsubsection{Full-duplex}

Wireless systems today are generally half-duplex, i.e., they can transmit or receive, but not both simultaneously. Full-duplex systems, however, are able to transmit and receive simultaneously. The challenge to design full-duplex systems is to reduce self-interference, which is due to signal received from a local transmitting antenna is usually much stronger than signal received from other nodes \cite{mobicom_11}. The main methods to reduce self-interference include analog cancelation techniques, digital cancelation, and antenna placement. The analog cancelation techniques treat self-interference as noise, and use noise canceling chips to reduce self-interference \cite{noise_chip}. The digital cancelation subtracts self-interference in the digital domain after ADC, and the antenna placement techniques place another TX antenna such that two transmit signals interfere destructively at RX antenna \cite{antenna_place}. Jain \emph{et al.} \cite{mobicom_11} use the balanced/unbalanced (balun) transformer to support wideband and high power systems. With the half-duplex constraint removed, the medium access control (MAC) needs to be redesigned to fully exploit the benefits of full-duplex. There are two transmission modes for the full-duplex systems, the bidirectional full-duplexing and the relay full-duplexing \cite{full_mode}. It is shown in \cite{tradeoff_full} that there is trade-off between spatial reuse and the full-duplex gain for traditional CSMA-style MAC, and the full-duplex gain is well below 2 in common cases from the network-level capacity gain view.

\begin{figure}
  \includegraphics[width=0.5\textwidth]{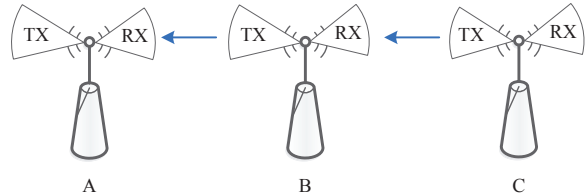}
\caption{The relay full-duplexing transmission by beamforming in the backhaul network.}
\label{fig:beam_full}       
\end{figure}

However, the current research on the full-duplex systems mainly focuses on the omnidirectional communications in the low frequency bands. There are several challenges to design full-duplex systems in the mmWave bands. First, the current full-duplex systems have the bandwidth at the order of 40MHz \cite{mobicom_11}, and the practical implementation of the full-duplex systems for the mmWave communications with a bandwidth of several GHz should be investigated and demonstrated. Second, since mmWave communications are inherently directional, the directional full-duplex systems with directional transmission and reception need to be developed. Miura \emph{et al.} \cite{directional_full} proposed a full-duplex node architecture with directional transmission and omnidirectional reception based on the full-duplex architecture in \cite{mobicom_11}.
With directional antennas used for both transmission and reception in the mmWave bands, the node architecture should be redesigned to take the characteristics of mmWave communications into account. The intuitive approach is to exploit the beamforming of transmission antennas and reception antennas to reduce self-interference. With the directions of transmit and receive antennas not directed towards each other, the self-interference can be reduced by beamforming. As shown in Fig. \ref{fig:beam_full}, with the beams of transmit and receive antennas of each AP directed in the inverse directions, the relay full-duplex transmission can be realized by beamforming in the line-type backhaul network since any transmitter is outside the receive range of the other receiver, where we assume AP A receives negligible interference from AP C.
Third, in the directional transmission scenario, the relationship between the spatial reuse and the full-duplex gain should be further investigated for the design of mmWave full-duplex networks in 5G cellular networks.

\subsection{Software Defined Architecture}

To overcome the challenges posed by mmWave communications, different
approaches from the physical layer to the network layer can be
exploited. However, every approach has its advantages and
shortcomings, and is efficient only in certain circumstances. We
should combine them intelligently to optimize the network
performance. On the other hand, with multiple APs deployed,
coordination among APs must be explicitly considered to achieve
goals such as efficient interference management and spatial reuse,
robust network connectivity, optimized load balance, and flexible
QoS guarantee.

Distributed network control does not scale well \cite{SoftRAN}, and
the latency will increase significantly as more APs are deployed.
Distributed control usually also has high control overhead.
Moreover, it is difficult to achieve intelligent control mechanisms
required for the complicate operational environments that involve
dynamic behaviors of accessing users and temporal variations of the
communication links in a distributed way. Therefore, we need
centralized and cross-layer control mechanisms to fully exploit the
potential of mmWave communications. Borrowing the idea of
Software-Defined Network (SDN) \cite{SDN1,SDN2}, which advocates
separating the control plane and data plane, and abstracting the
control functions of the network into a logically centralized
controller to expose the functions deeply hidden inside the network
stack to higher layers, the software defined network will be a
promising architecture for mmWave communication systems to achieve
flexible and intelligent network control. To deploy the software
defined mmWave communication system, the interface between the
control plane and data plane (e.g., OpenFlow \cite{OpenFlow}) should
be designed elaborately to facilitate the efficient and centralized
control of the control plane. Besides, efficient and centralized
control mechanisms of the control plane, and network state
information measurements, on which control mechanisms are based, are
also open problems which warrant further investigations.

\subsection{Control Mechanisms}

 To achieve high network performance, effective and efficient mechanisms on interference management, transmission scheduling,
 mobility management and handover, beamforming (BF), and anti-blockage need to be further
 investigated.

\begin{table*}
\centering
\newcommand{\tabincell}[2]{\begin{tabular}{@{}#1@{}}#2\end{tabular}}
\caption{Comparison of MAC protocols for mmWave communications}
\label{tab:2}
\begin{tabular}{r@{}lr@{}lr@{}lr@{}lr@{}l} \hline
\multicolumn{2}{c}{} &&TDMA-based &&spatial reuse && anti-blockage
&& \tabincell{l}{centralized or \\distributed}\\ \hline
\multicolumn{2}{l}{Directional CSMA/CA \cite{Gong}} && \tabincell{l}{No} && \tabincell{l}{not specified}  && not specified  && centralized  \\
\multicolumn{2}{l}{MRDMAC \cite{MRDMAC}}       && \tabincell{l}{No} && \tabincell{l}{not specified, \\for flows \\between the AP \\and WTs}     && \tabincell{l}{supported, \\by multi-hop \\relaying}  && centralized \\
\multicolumn{2}{l}{MDMAC \cite{MDMAC}}     && \tabincell{l}{No, time division \\multiplexing (TDM)} && \tabincell{l}{not supported, \\to achieve \\approximate \\TDM schedules}      && not specified  && distributed  \\
\multicolumn{2}{l}{FDMAC \cite{mao}}     && \tabincell{l}{No, frame-based} && \tabincell{l}{supported, by \\greedy coloring \\(GC) algorithm} && not specified && centralized \\
\multicolumn{2}{l}{D-CoopMAC \cite{chenqian}}     &&\tabincell{l}{No, based on \\IEEE 802.11ad} && \tabincell{l}{not specified}      && not specified   && centralized  \\
\multicolumn{2}{l}{REX \cite{EX Region}}   && \tabincell{l}{Yes, based on \\IEEE 802.15.3} && \tabincell{l}{supported, by a \\randomized ER \\based scheduling \\scheme}      && not specified && centralized  \\
\multicolumn{2}{l}{Spatial sharing \cite{chen_2}} &&\tabincell{l}{No, based on\\ IEEE 802.11ad} && \tabincell{l}{supported, \\based on the \\BF information}      && not specified && centralized  \\
\multicolumn{2}{l}{MHCT \cite{Qiao_7}}   && \tabincell{l}{Yes, based on \\IEEE 802.15.3c} && \tabincell{l}{supported, by \\the MHCT scheme} && \tabincell{l}{supported, \\by multi-hop \\relaying} && centralized  \\
\multicolumn{2}{l}{STDMA based scheduling \cite{Qiao}}   && \tabincell{l}{Yes, based on \\IEEE 802.15.3c} && \tabincell{l}{supported, \\by concurrent \\transmission \\scheduling \\algorithm} && not specified && centralized  \\
\multicolumn{2}{l}{VTSA \cite{Qiao_6}}   && \tabincell{l}{Yes, based on \\IEEE 802.15.3c} && \tabincell{l}{supported} && not specified && centralized  \\
\multicolumn{2}{l}{Spatial reuse TDMA \cite{mao_12}}   && \tabincell{l}{Yes, based on \\IEEE
802.15.3c} && \tabincell{l}{supported} && not specified && centralized  \\
\multicolumn{2}{l}{DtDMAC \cite{DtDMAC}}   && \tabincell{l}{No, an exponential \\backoff procedure \\for asynchronous \\operation} && \tabincell{l}{supported} && not specified && distributed  \\
\multicolumn{2}{l}{BRDMAC \cite{tvt_own}}   && \tabincell{l}{No, frame-based} && \tabincell{l}{supported, \\ by the SINR \\based concurrent \\transmission \\scheduling} && \tabincell{l}{supported, \\by the two-hop \\relaying} && centralized  \\
\multicolumn{2}{l}{CTA-PSO \cite{2_8}}   && \tabincell{l}{No, based on \\IEEE 802.11ad} && \tabincell{l}{not specified} && \tabincell{l}{supported, \\by the link switch \\relay method} && centralized  \\
\hline
\end{tabular}
\end{table*}



 In table \ref{tab:2}, we compare several typical MAC
 protocols for mmWave networks in terms of some key aspects. From the table, we can
 observe that every protocol has its advantages and shortcomings,
 and more efficient and robust protocols need to be developed to exploit the potential of spatial reuse and also overcome
 blockage. To achieve better network performance, centralized
 protocols are preferred. On the other hand, most work focuses on the scenario of one BSS
\cite{Qiao} and does not consider
 the interference among different BSSs.

\begin{table*}
\centering
\caption{The anti-blockage works classification}
\label{tab:blockage}       
\begin{tabular}{lcc}
\hline\noalign{\smallskip}
Anti-blockage Strategies & References & Layer\\
\noalign{\smallskip}\hline\noalign{\smallskip}
Reflection or NLOS transmission & \cite{robust,2_6,beam switching,EG,Xiao} & PHY/MAC\\
Relaying & \cite{MRDMAC,blockage and directivity
conference,Lan,2_11,tvt_own}& MAC \\
Multi-AP diversity & \cite{multiap}& MAC \\
multipath routing & \cite{multipath}& Routing \\
\noalign{\smallskip}\hline
\end{tabular}
\end{table*}

 As regards anti-blockage, we classify the works for anti-blockage according to the strategies adopted, and the layers where the strategies are performed in table \ref{tab:blockage}. Although there are several approaches such as beam switching from a LOS path to a NLOS path
 \cite{beam switching}, relaying \cite{MRDMAC}, performing handovers between APs
 \cite{multiap}, and exploiting spatial diversity \cite{EG}, multipath routing \cite{multipath}, each of
 them has limitations, and it is efficient only under certain
 conditions. For example, performing handovers between APs is
 effective only when there is a direct path between the user device
 and another neighboring AP. Beam switching to a NLOS path is usually a good choice \cite{beam
switching}. In
 some cases, however, the NLOS path is difficult to find, or for a high-rate flow such as HDTV,
 the transmission rate supported by the NLOS path cannot meet the throughput
 requirement. In this case, relaying may be a good choice if the links in the
 relay path have high channel quality. Exploiting spatial diversity
 increases complexity and can not guarantee the quality of service
 (QoS). Therefore, how to combine these approaches and apply them appropriately in
order to ensure robust network connectivity and improve network
performance remains an open problem which warrants further
investigations.

On the other hand, due to the diverse applications as well as the
complexity and variability of the indoor environment, mmWave
communication systems should be able to adapt to different traffic
patterns and time-varying channel states. Even though a variety of
approaches or protocols have been proposed, each optimized for a
specific application or channel state, it is essential to be able to
switch among them appropriately and intelligently, or to combine
them efficiently, according to the actual network state. Open
problems on dynamical adaption include how to combine TDMA and
CSMA/CA intelligently to adapt to a variety of applications, how to
select the MCSs according to the time-varying channel states, and
how to manage user mobility to improve the network performance.

\subsection{Network State Measurement}

Efficient and intelligent network control are based on accurate and
comprehensive network state information obtained by efficient
measurement mechanisms. There already exist some work on the
measurement of network state information. Ning \emph{et al.}
\cite{bootstrapping} considered the process of neighbor discovery in
60 GHz indoor wireless networks, and examine direct discovery and
gossip-based discovery, from which the up-to-date network topology
or node location information can be obtained. Kim \emph{et al.}
\cite{D-ND} analyzed the directional neighbor discovery process
based on the IEEE 802.15.3c standard. Park \emph{et al.} \cite{2_9} proposed a multi-band directional neighbor discovery scheme, where management procedures are carried out in the 2.4 GHz band with the omni-directional antennas while data transmissions are performed in the 60 GHz band with
directional antennas, to reduce the neighbor discovery time and energy consumption. Chen \emph{et al.}
\cite{chen_2} designed a beamforming (BF) information table that
records all the beamforming training results among clients that can
be established at the AP.

However, most of the current work focuses on the network state
information measurement within one BSS. For user devices in the
overlapped region of two BSSs, the link quality information and BF
information between the devices and the neighbouring APs are also
required to perform handover and interference management. To
maximize concurrent transmissions among different BSSs, the
interference between different BSSs must be estimated as accurate as
possible. On the other hand, to ensure the real-time network
control, all the network state measurements should be completed
within the shortest possible time. Thus, efficient measurement
mechanisms are open problems, which need to be extensively
investigated to facilitate the deployment of mmWave communication
systems in the future.

\subsection{Heterogeneous Networking}

\begin{figure}
  \includegraphics[width=0.5\textwidth]{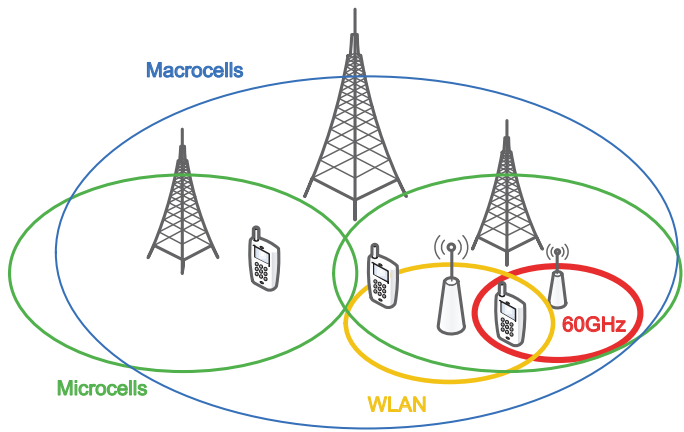}
\caption{Heterogeneous networks consisting of macrocells,
microcells, WLANs, and picocells in the 60 GHz band}
\label{fig:1}       
\end{figure}

Due to the limited coverage area of mmWave communications, mmWave
communication systems need to coexist with other systems of other
bands, such as LTE and WiFi. Therefore, mmWave networks will be
inherently heterogeneous \cite{mmW-cellular,2_24}. As shown in Fig.
\ref{fig:1}, short-range picocells in the 60 GHz band will coexist
with macrocells and microcells in other bands \cite{heterogeneous
cells}. We can observe that cells in the microwave bands have larger
coverage areas, while smaller cells such as BSSs in the 60 GHz band
have higher capacity.

Heterogeneous networking has gained considerable attention from
academia and industry \cite{heterogeneous networking 1,heterogeneous
networking 2,heterogeneous networking 3,heterogeneous networking 4}. Mehrpouyan \emph{et al.} \cite{HetNets} introduced a
novel millimeter-wave HetNet paradigm, termed
hybrid HetNet, which exploits the vast bandwidth
and propagation characteristics in the 60 GHz
and 70--80 GHz bands to reduce the impact of
interference in HetNets. In heterogeneous networks, interaction and cooperation between
different kinds of networks become the key to explore the potential
of heterogeneous networking to solve the problems of mobility
management, vertical handover, mobile data offloading from
macrocells to microcells \cite{mobile data offloading}, inter-cell
interference management, etc. Qiao \emph{et al.} \cite{2_29} introduced
an mmWave+4G system architecture with TDMA-based MAC structure as a candidate for
5G cellular networks, where the control functions are performed in the 4G system. The high capacity of mmWave
communications can offload traffic from the macrocells and provide
better services for traffic with high throughput requirements. On
the other hand, handovers between base stations (BS) of macrocells
and APs in the mmWave band are able to address problems such as
blockage, mobility management, load balancing, etc. For mmWave
networks, there are also studies advocating distributing the control
messages for channel access and coordination both on mmWave and
microwave bands \cite{MMB}. Thus, a part of important control
signals such as synchronization or channel access requests can be
transmitted in the microwave band omni-directionally. Thus, network
coupling, the level of integration between different networks, has
an important impact on the system performance \cite{heterogeneous
book}. Tight coupling is beneficial to achieve better performance,
while loose coupling has low complexity. Therefore, there is a
tradeoff between complexity and performance in heterogeneous
networking. Meanwhile, software defined architecture with flexible
programmability, such as OpenRadio \cite{OpenRadio}, is a promising
candidate to achieve tight coupling between networks.

\section{Conclusions}\label{sec:5}


With the potential to offer orders of magnitude greater capacity
over current communication systems, mmWave communications become a promising candidate for the 5G mobile
networks. In this paper, we carry out a survey of mmWave communications for 5G. The characteristics of mmWave communications promote the
redesign of architectures and protocols to address the challenges,
including integrated circuits and system design, interference management and spatial reuse, anti-blockage,
and dynamics due to mobility. The current solutions have been
overviewed and compared in terms of effectiveness, efficiency, and
complexity. The potential applications of mmWave communications in 5G are also discussed. Open research issues, related to the new physical technology, the software defined
architecture, the measurements of network state information, the efficient
control mechanisms, and the heterogeneous networking, have been
discussed to promote the development of mmWave communications in 5G.




\end{spacing}
\end{document}